\pdfoutput=1
\documentclass[a4paper,onecolumn,unpublished,nopdfoutputerror]{quantumarticle}
\usepackage[T1]{fontenc}
\usepackage[utf8]{inputenc}
\usepackage{graphicx}
\usepackage{mathrsfs}
\usepackage{amsmath,amssymb,amsthm,bm}
\usepackage{xcolor}
\usepackage[colorlinks=true,linkcolor=purple,citecolor=purple,urlcolor=blue]{hyperref}
\usepackage[nameinlink,capitalize]{cleveref}

\newtheorem{proposition}{Proposition}
\newtheorem{lemma}{Lemma}

\theoremstyle{definition}
\newtheorem{remark}{Remark}

\DeclareMathOperator{\Tr}{Tr}
\DeclareMathOperator{\Ran}{Ran}
\DeclareMathOperator{\rank}{rank}

\newcommand{\ket}[1]{|#1\rangle}
\newcommand{\bra}[1]{\langle #1|}

\newcommand{\id}{\mathrm{id}}
\newcommand{\SN}{\mathrm{SN}}

\begin{document}

\title{Competing coherence geometries in positive maps: positivity
hierarchies, atomicity, and optimal entanglement
  witnesses}
\author{Massimiliano F. Sacchi}
\orcid{0000-0002-8909-2196}
\affiliation{Consiglio Nazionale delle Ricerche, Istituto di Fotonica e Nanotecnologie, Piazza Leonardo da Vinci 32, I-20133, Milano, Italy}
\affiliation{Dipartimento di Fisica, Universit\`a degli Studi di Pavia, Via Agostino Bassi 6, I-27100, Pavia, Italy}

\begin{abstract}
We introduce a family of positive maps on qudits whose Choi operators
combine two competing coherence patterns: independent exchange
couplings and a collective set of diagonal couplings. For every mixed
member of the family, we derive exact conditions for positivity,
complete positivity, and all intermediate levels of $k$-positivity,
and we determine the complete phase structure of decomposability and
atomicity. The higher-positivity thresholds depend on the channel
multiplicities, whereas the exact decomposability boundary is
universal in scale-free coordinates. Its proof combines an explicit
positive-plus-copositive decomposition with a matching family of
positive-partial-transpose (PPT) states. The latter form PPT edge
states whose detection region is an ellipse, or a disk after
collective rescaling. An embedded-qutrit certificate shows that
indecomposability and atomicity coincide throughout the mixed
family. In contrast, the purely collective endpoint is decomposable
over its entire positive interval. Finally, at maximal coherence, an
explicit positive subtraction produces an atomic optimal entanglement
witness with the spanning property and a strictly larger PPT-detection
set. The study thus provides an exactly solvable setting in which
$k$-positivity, decomposability, atomicity, PPT-edge-state geometry,
and witness optimality can be compared in arbitrary local dimension.
\end{abstract}

\maketitle
\section{Introduction}
\label{sec:intro}
Entanglement is a defining nonclassical feature of quantum theory and
a central resource for quantum information processing. Deciding
whether a mixed bipartite state is separable remains
difficult. Positive maps provide a complete, although generally
intractable, formulation of the problem: a state is separable if and
only if it remains positive under the ampliation
$\mathrm{id}\otimes\Phi$ for every positive linear map $\Phi$
\cite{Horodecki1996,Horodecki2009}. Through the Choi--Jamio\l{}kowski
correspondence, the same question becomes one of block positivity for
bipartite operators, thereby connecting the separability problem with
the geometry of cones of maps, states, and entanglement witnesses
\cite{Jamiolkowski1972,Choi1975,Terhal2000,ChruscinskiSarbicki2014}.

Indecomposable positive maps are especially important for entanglement
theory. The positive-partial-transpose (PPT) criterion is necessary
for separability, but ceases to be sufficient beyond $\mathbb
C^2\otimes\mathbb C^2$ and $\mathbb C^2\otimes\mathbb C^3$ bipartite
states; decomposable maps cannot detect PPT entangled states, whereas
every indecomposable witness detects at least one such state
\cite{Peres1996,Horodecki1996,Lewenstein2000,HaKyePark2003}. The
low-dimensional decomposability results of St\o rmer and Woronowicz
\cite{Stormer1963,Woronowicz1976} contrast sharply with the richer
geometry beginning in $\mathbb C^3\otimes\mathbb C^3$, where the Choi
map and its generalizations provide prototypical indecomposable
examples
\cite{Choi1972,Choi1975,ChoKyeLee1992,Stormer2013}. Subsequent
analyses of Choi-type maps clarified positivity, optimality,
extremality, exposedness, and the associated PPT faces
\cite{HaKye2004,HaKye2005,HaKyeExposed2011,HaKye2012,ChruscinskiSarbicki2012}.

A broad range of constructions is available. Positive maps have been
obtained from tensor-cone duality, spectral data, orthogonal matrices,
positive biquadratic forms, elementary operators, and merging procedures
\cite{MajewskiMarciniak2001,Kossakowski2003,ChruscinskiKossakowski2007,
ChruscinskiKossakowski2009,SkowronekZyczkowski2009,Hou2010,
MarciniakRutkowski2017}. Questions of decomposability and persistence under
tensor powers lead to further structural difficulties
\cite{MullerHermes2018}. Related symmetry-based approaches reduce positive-map
conditions to structured matrices and connect them with copositivity and
extendibility hierarchies \cite{SinghNechita2021,GulatiNechitaPark2025}.
Recent work has also emphasized experimentally accessible criteria based on
symmetric measurements and moments of positive maps, symmetry-based
distillation tests, and differentiable semidefinite programming for the
numerical generation of nondecomposable maps
\cite{LiYaoFeiFanMa2024,MallickMaityGangulyMajumdar2025,Rico2026,arx}.
These developments illustrate both the variety of known examples and the
persistent difficulty of determining positivity, $k$-positivity,
decomposability, atomicity, and optimality within a single analytic model.

The distinction among these properties is essential. The $k$-positive
hierarchy refines entanglement detection through Schmidt number
\cite{SanperaBrussLewenstein2001,HorodeckiLewensteinVidalCirac2000},
while atomic maps can detect the weakest form of quantum entanglement,
namely that of PPT states whose state and partial transpose both
have Schmidt number two
\cite{Lewenstein2000,SanperaBrussLewenstein2001,ChruscinskiSarbicki2014}. 
Optimal witnesses address a different question, namely whether the detected
set can be enlarged by subtracting a positive operator
\cite{Lewenstein2000,HaKye2012,ChruscinskiSarbicki2014}. On the state side,
edge states and the facial structures of the separable and PPT cones provide
the natural dual geometric objects
\cite{Horodecki1997,Kye2011Faces,FacialStructures2012,HaKyePark2003}.
Analytic families that keep these notions visible at the same time are
therefore particularly valuable.

The starting point of the present work is the sparse qutrit family studied in
Ref.~\cite{CompanionQutrit}. In that model, two independently weighted
coherence channels in the Choi operator lead to an exact
positivity square, a circular decomposability boundary,
explicitly adapted PPT edge states, and an optimal refinement of the
maximally coherent witness. A complementary qutrit construction, with exact
nondecomposability thresholds and PPT-entanglement transitions, was analyzed
in Ref.~\cite{PoderiniMorgillo2026}. Extending the present construction to
arbitrary local dimension is not unique and is not achieved by simply adjoining
inactive basis states. Additional directions can be arranged either
as independent exchange couplings
$\ket{1i}\!\bra{i1}+\ket{i1}\!\bra{1i}$ or as collective diagonal
couplings $\ket{11}\!\bra{jj}+\ket{jj}\!\bra{11}$. We refer to the latter as collective coherences because all such couplings
share the central vector $\ket{11}$; on the diagonal subspace, they form a
star-shaped connectivity pattern. The two patterns behave differently under
complete and higher positivity, and the limiting family containing
only collective couplings has no counterpart in the original
two-parameter qutrit case.

The present construction is an arbitrary-dimensional extension of the
coherence-anisotropic qutrit family, but it also exhibits
genuinely higher-dimensional behavior. It is complementary to the
diagonal-anisotropic qudit families studied in a concurrent work, where
the independent parameters enter the diagonal sectors rather than the
coherence amplitudes \cite{diag}. The competition between exchange and
collective coherences reveals three higher-dimensional effects: a
multiplicity-dependent hierarchy of higher positivity, a
decomposability boundary independent of the channel multiplicities,
and a qualitatively different purely collective endpoint.

The sparse support makes this simultaneous classification analytically
tractable without trivializing the resulting cone geometry.  The
product-state positivity condition reduces to a one-variable
optimization, in which the only relevant parameter is the fraction of
the input norm carried by one distinguished basis component. Complete
and $k$-positivity are controlled by small active sectors.
Decomposability is established by an explicit positive-plus-copositive
decomposition, while a matching family of PPT operators proves the
converse implication. These PPT operators are edge states, and their
detection regions become ellipses whose axes directly record the two
channel multiplicities. Finally, an embedded qutrit subspace supplies
PPT states of Schmidt number two for both the state and its partial
transpose, proving that indecomposability and atomicity coincide
throughout the mixed family.

Our main results give the complete phase structure generated by the competing
coherences. For every mixed member of the family, we determine the exact
positivity and $k$-positivity regions, the necessary and sufficient condition
for decomposability, and the complementary atomic region. We construct
full-space PPT edge states and determine their detection geometry. At maximal
coherence, an explicit positive subtraction yields an optimal witness with a
strictly larger PPT-detection set. The purely collective endpoint is
classified separately and is decomposable throughout its positive region.

The remainder of the paper is organized as follows.
Section~\ref{sec:prelim} fixes the terminology and notation.
Section~\ref{sec:family} introduces the family of sparse maps and
determines its positivity, complete-positivity, and $k$-positivity
regions. Section~\ref{sec:decomposability} establishes the exact
decomposability and atomicity boundaries, treats the purely collective
endpoint, and collects the complete phase
diagram. Section~\ref{sec:edge-states} develops the associated PPT
edge states and their detection
geometry. Section~\ref{sec:optimal-refinement} constructs and analyzes
the optimal refinement at maximal coherence.
Section~\ref{sec:conclusion} concludes the paper.

\section{Preliminaries and notation}
\label{sec:prelim}
We collect the standard notions and notation used throughout the paper; see
Refs.~\cite{Stormer2013,Horodecki2009,ChruscinskiSarbicki2014} for
comprehensive accounts. Let $M_d$ denote the algebra of complex
$d\times d$ matrices, and let $\Phi:M_d\to M_d$ be a
Hermiticity-preserving linear map. Its Choi operator is given by
\begin{equation}
 C_\Phi=\sum_{i,j=1}^d\ket i\bra j\otimes\Phi(\ket i\bra j).
\label{eq:choi-definition}
\end{equation} The map $\Phi$ is positive if $\Phi(X)\ge0$ for every
$X\ge0$, $k$-positive if $\id_k\otimes\Phi$ is positive, where
$\id_k:M_k\to M_k$ is the identity map, and completely
positive if it is $k$-positive for every $k$. Under the
Choi--Jamio\l{}kowski correspondence, these conditions are equivalent,
respectively, to block positivity of $C_\Phi$, nonnegativity of $C_\Phi$ on
vectors of Schmidt rank at most $k$, and $C_\Phi\ge0$
\cite{Jamiolkowski1972,Choi1975,TerhalHorodecki2000}. Here, block positivity
means
\begin{equation}
 \bra{x,y}C_\Phi\ket{x,y}\ge0
 \qquad\text{for all }\ket x,\ket y\in\mathbb C^d.
\end{equation}

Partial transposition on the second subsystem in the basis introduced
in Eq.~\eqref{eq:choi-definition} is denoted by $\Gamma$. A map is
$k$-copositive if the composition of $\Phi$ with transposition is
$k$-positive; equivalently, $C_\Phi^\Gamma$ is nonnegative on all
vectors of Schmidt rank at most $k$. Complete copositivity is defined
analogously. A positive map, or its Choi witness $W=C_\Phi$, is
decomposable if
\begin{equation}
 W=P+Q^\Gamma,\qquad P,Q\ge0.
\label{eq:decomposable-form}
\end{equation}
It is atomic if it cannot be written as the sum of a $2$-positive and a
$2$-copositive map. Equivalently, its Choi operator cannot be written as
\begin{equation}
\label{eq:2decomp}
 W=W_1+W_2^\Gamma,
\end{equation}
where $W_1$ and $W_2$ are nonnegative on all vectors of Schmidt rank at most
two \cite{ChruscinskiSarbicki2014}.

The Schmidt number $\SN(\rho)$ of a positive operator $\rho$ is the smallest
integer $k$ for which $\rho$ is a finite sum of projectors onto vectors of
Schmidt rank at most $k$ \cite{TerhalHorodecki2000,SanperaBrussLewenstein2001}.
Consequently, a PPT state satisfying
\begin{equation}
 \SN(\rho)\le2,\qquad \SN(\rho^\Gamma)\le2,
 \qquad \Tr(W\rho)<0
\label{eq:atomicity-certificate}
\end{equation}
certifies atomicity: if Eq.~\eqref{eq:2decomp} held, then both contributions to
$\Tr(W\rho)$ would be nonnegative.

A PPT state $\rho$ is an \emph{edge state} if no nonzero product vector
$\ket{x,y}\in\Ran\rho$ satisfies
$\ket{x,\overline y}\in\Ran\rho^\Gamma$, where complex conjugation is taken in
the basis defining $\Gamma$. Equivalently, no positive multiple of a product
projector can be subtracted while preserving positivity and the PPT property
\cite{Horodecki1997,HorodeckiLewensteinVidalCirac2000,Lewenstein2000}.

Finally, an entanglement witness $W$ is \emph{optimal} if no nonzero positive operator
$D\ge0$ can be subtracted while preserving block positivity. Its zero product
vectors form the set
\begin{equation}
 \mathcal P_W:=\{\ket{x,y}:\bra{x,y}W\ket{x,y}=0\}.
\end{equation}
The witness has the \emph{spanning property} if
$\operatorname{span}\mathcal P_W=\mathbb C^d\otimes\mathbb C^d$. The spanning
property is sufficient, but not necessary, for optimality
\cite{Lewenstein2000,ChruscinskiSarbicki2014}.

\section{Sparse family and positivity hierarchy}
\label{sec:family}
We now introduce the family of maps along with its sparse Choi operators and then
classify its positivity hierarchy.
\subsection{Definition of the family} 
Fix $a>0$ and $d\geq3$. For $r=1,\ldots,d-1$, define the
Hermiticity-preserving map $\Phi^{(a)}_{d,r}(w,z):M_d\to M_d$ as
follows
\begin{align}
\Phi^{(a)}_{d,r}(w,z)[X]
&=a\Tr(X)I_d
+w
\sum_{i=2}^{d-r}
\bigl(x_{1i}\ket i\bra1+x_{i1}\ket1\bra i\bigr) +z
\sum_{j=d-r+1}^{d}
\bigl(x_{1j}\ket1\bra j+x_{j1}\ket j\bra1\bigr),
\label{eq:map-action}
\end{align}
where $X=(x_{mn})\in M_d$. Here and below, $I_n$ denotes the identity operator on an
$n$-dimensional space. We assume $w,z\ge0$ whenever the corresponding
parameter is present. The corresponding Choi operator
$W^{(a)}_{d,r}(w,z)=C_{\Phi^{(a)}_{d,r}(w,z)}$ is then given by 
\begin{equation}
\label{eq:unified-family}
W^{(a)}_{d,r}(w,z)=aI_{d^2}
+w\sum_{i=2}^{d-r}
\bigl(\ket{1i}\!\bra{i1}+\ket{i1}\!\bra{1i}\bigr)
+z\sum_{j=d-r+1}^{d}
\bigl(\ket{11}\!\bra{jj}+\ket{jj}\!\bra{11}\bigr).
\end{equation}
The Choi form in Eq.~\eqref{eq:unified-family} will be more convenient
for most of the analysis.

It is useful to introduce the index sets
\begin{equation}
 I:=\{2,\ldots,d-r\},\qquad
 J:=\{d-r+1,\ldots,d\},
\end{equation}
and their cardinalities
\begin{equation}
 p:=|I|=d-r-1,\qquad |J|=r.
\end{equation}
Thus the first sum in Eq.~\eqref{eq:unified-family} contains $p$ independent
exchange coherences, whereas the second contains $r$ collective coherences
sharing $\ket{11}$. For $r=1$ the construction
reduces, at $d=3$, to the anisotropic qutrit family of
Ref.~\cite{CompanionQutrit}. Intermediate values $1<r<d-1$ contain both
channel types. At $r=d-1$ the exchange sector disappears and one obtains
\begin{equation}
\label{eq:tilde-family}
\widetilde W^{(a)}_d(z)=aI_{d^2}
+z\sum_{j=2}^{d}\bigl(\ket{11}\!\bra{jj}+\ket{jj}\!\bra{11}\bigr).
\end{equation}

As a concrete illustration, consider $a=1$, $d=5$, and $r=2$. Then
$I=\{2,3\}$ and $J=\{4,5\}$, so
\begin{equation}
\begin{split}
 W^{(1)}_{5,2}(w,z)=I_{25}
 &+w\bigl(\ket{12}\!\bra{21}+\ket{21}\!\bra{12}
     +\ket{13}\!\bra{31}+\ket{31}\!\bra{13}\bigr)\\
 &+z\bigl(\ket{11}\!\bra{44}+\ket{44}\!\bra{11}
     +\ket{11}\!\bra{55}+\ket{55}\!\bra{11}\bigr).
\end{split}
\label{eq:d5-r2-example}
\end{equation}

In the ordered active basis
$\{\ket{11},\ket{44},\ket{55},\ket{12},\ket{21},\ket{13},\ket{31}\}$,
the operator is the direct sum of $I_{18}$ and the sparse matrix
\begin{equation}
\label{eq:d5-r2-matrix}
\begin{pmatrix}
1&z&z&0&0&0&0\\
z&1&0&0&0&0&0\\
z&0&1&0&0&0&0\\
0&0&0&1&w&0&0\\
0&0&0&w&1&0&0\\
0&0&0&0&0&1&w\\
0&0&0&0&0&w&1
\end{pmatrix}.
\end{equation}
The matrix displays two independent $2\times2$ exchange blocks and one
three-dimensional collective block. These two structures underlie the phase
structure analyzed below.

\subsection{Positivity and complete positivity}
\label{sec:phase}
The example in Eqs.~\eqref{eq:d5-r2-example} and
\eqref{eq:d5-r2-matrix} separates the two mechanisms relevant to positivity:
independent exchange blocks and one collective block. This decomposition
leads directly to exact criteria. Block positivity follows from an
optimization over product vectors, whereas complete positivity is obtained
from the spectra of the active blocks.

\begin{proposition}[Block positivity]
\label{prop:positivity}
If $r< d-1$, then $W^{(a)}_{d,r}(w,z)$ is block positive if and only if
\begin{equation}
0\le w\le2a,
\qquad
0\le z\le2a.
\end{equation}
If $r=d-1$, then $\widetilde W^{(a)}_d(z)$ is block positive if and only if
\begin{equation}
0\le z\le2a.
\end{equation}
\end{proposition}

\begin{proof}
Let $\ket{x}=\sum_{i=1}^d x_i\ket i$ and $\ket{y}=\sum_{i=1}^d y_i\ket i$ be unit
vectors, and set $\mu_i:=|x_i|\,|y_i|$. By Cauchy--Schwarz,
\begin{equation}
 \mu_i\ge0,\qquad \sum_{i=1}^d\mu_i\le1.
\end{equation}
The expectation on product vectors is
\begin{equation}
\bra{x,y}W^{(a)}_{d,r}(w,z)\ket{x,y}
=a+2w\sum_{i\in I}\Re(\overline{x_1}x_i y_1\overline{y_i})
\quad+2z\sum_{j\in J}\Re(\overline{x_1}\,\overline{y_1}x_jy_j).
\end{equation}
For fixed moduli, each real part is bounded below by
$-|x_1||y_1||x_i||y_i|=-\mu_1\mu_i$. The phases of the components with
indices in $I$ and $J$ can be chosen independently, so all these lower bounds
are attained simultaneously. Hence the minimum over phases is
\begin{equation}
 a-2\mu_1\left(w\sum_{i\in I}\mu_i+z\sum_{j\in J}\mu_j\right).
\label{eq:block-positive-envelope}
\end{equation}
For fixed $\mu_1=s$, the weighted sum in parentheses is at most
$\max\{w,z\}(1-s)$; equality is obtained by concentrating the remaining
weight in an index belonging to the channel of maximal strength. Therefore the minimum of the product expectation in
Eq.~\eqref{eq:block-positive-envelope} is
\begin{equation}
 a-2s(1-s)\max\{w,z\}.
\end{equation}
Since $s(1-s)\le1/4$, block positivity is equivalent to
$\max\{w,z\}\le2a$ when both channel classes are present. If $p=0$, only the
$z$ term remains and the condition reduces to $z\le2a$.
\end{proof}

\begin{proposition}[Complete positivity]
\label{prop:cp}
The Choi operator $W^{(a)}_{d,r}(w,z)$ is positive semidefinite if and only if
\begin{equation}
\label{eq:cp-criterion}
\begin{cases}
0\le w\le a, & r<d-1,\\[1mm]
0\le z\le \dfrac{a}{\sqrt r}, & 1\le r\le d-1.
\end{cases}
\end{equation}
In particular,
\begin{equation}
\widetilde W^{(a)}_d(z)\ge0
\quad\Longleftrightarrow\quad
0\le z\le\frac{a}{\sqrt{d-1}}.
\end{equation}
\end{proposition}

\begin{proof}
After permuting the computational basis, each exchange pair
$\{\ket{1i},\ket{i1}\}$, $i\in I$, gives the block
\begin{equation}
 \begin{pmatrix}a&w\\ w&a\end{pmatrix},
\end{equation}
whose eigenvalues are $a\pm w$. The collective sector is supported on
$\operatorname{span}\{\ket{11},\ket{jj}:j\in J\}$, which has dimension
$r+1$. Define the normalized collective vector
\begin{equation}
 \ket{s_J}:=\frac{1}{\sqrt r}\sum_{j\in J}\ket{jj}.
\end{equation}
The collective sector then decomposes orthogonally as
\begin{equation}
 \operatorname{span}\{\ket{11},\ket{jj}:j\in J\}
 =\operatorname{span}\{\ket{11},\ket{s_J}\}\oplus\mathcal S_J^\perp,
\end{equation}
where
\begin{equation}
 \mathcal S_J^\perp
 :=\left\{\sum_{j\in J}c_j\ket{jj}:\sum_{j\in J}c_j=0\right\}
\end{equation}
has dimension $r-1$. For every $\ket\eta\in\mathcal S_J^\perp$, the
collective coherence term vanishes, so
\begin{equation}
 W^{(a)}_{d,r}(w,z)\ket\eta=a\ket\eta.
\end{equation}
On the remaining two-dimensional subspace, the ordered basis
$\{\ket{11},\ket{s_J}\}$ gives
\begin{equation}
 \left.W^{(a)}_{d,r}(w,z)\right|_{\operatorname{span}\{\ket{11},\ket{s_J}\}}
 =\begin{pmatrix}a&z\sqrt r\\ z\sqrt r&a\end{pmatrix},
\end{equation}
whose eigenvalues are $a\pm z\sqrt r$. All remaining basis vectors carry
the eigenvalue $a$. The stated conditions follow by requiring that all
eigenvalues are nonnegative.
\end{proof}

\subsection{Higher positivity}
\label{sec:higher-positivity}
We now determine the intermediate levels between positivity and complete
positivity. Every exchange coherence is already probed by a Schmidt-rank-two
antisymmetric vector. The collective sector behaves differently: a test vector
supported on $\ket{11}$ and on $\ell$ independent diagonal vectors
$\ket{j_qj_q}$ has Schmidt rank $\ell+1$. This accounts for the factor
$\min\{r,k-1\}$ in the exact threshold below.

We first prove the only estimate needed for the collective sector. For a matrix
$C$, let $\|C\|_{\mathrm F}$ denote its Frobenius norm, and let
$s_1(C)\ge s_2(C)\ge\cdots$ denote its singular values.

\begin{lemma}[Collective coherence bound]
\label{lem:collective-coherence}
Let $C=(c_{pq})\in M_{r+1}$ satisfy $\rank C\le k$, with $k\ge2$, and set
\begin{equation}
 m:=\min\{r,k-1\}.
\end{equation}
If the first index is distinguished and the other $r$ indices form the set
$J$, then
\begin{equation}
 2\left|\overline{c_{11}}\sum_{j\in J}c_{jj}\right|
 \le \sqrt m\,\|C\|_{\mathrm F}^2.
\label{eq:collective-coherence-bound}
\end{equation}
The constant $\sqrt m$ is optimal.
\end{lemma}

\begin{proof}
We use the following standard consequence of von Neumann's trace inequality.
If $\rank C\le k$, then for every matrix $A$ of the same size,
\begin{equation}
 |\Tr(A^\dagger C)|
 \le \sum_{q=1}^{k}s_q(A)s_q(C)
 \le
 \left(\sum_{q=1}^{k}s_q(A)^2\right)^{1/2}
 \|C\|_{\mathrm F}.
\label{eq:rank-constrained-trace-bound}
\end{equation}
Put
\begin{equation}
 \alpha:=|c_{11}|,
 \qquad
 \beta:=\left|\sum_{j\in J}c_{jj}\right|,
\end{equation}
and choose phases $\theta_1,\theta_2$ such that
\begin{equation}
 e^{-i\theta_1}c_{11}=\alpha,
 \qquad
 e^{-i\theta_2}\sum_{j\in J}c_{jj}=\beta.
\end{equation}
For $t>0$, define
\begin{equation}
 A_t:=t e^{i\theta_1}\ket1\!\bra1
 +t^{-1}e^{i\theta_2}P_J,
 \qquad
 P_J:=\sum_{j\in J}\ket j\!\bra j.
\end{equation}
Then
\begin{equation}
 |\Tr(A_t^\dagger C)|=t\alpha+t^{-1}\beta.
\label{eq:At-pairing}
\end{equation}
Choose $t=m^{1/4}$. The singular values of $A_t$ are $t$ once and
$t^{-1}$ with multiplicity $r$. Since $t\ge1$, the sum of the squares of the
$k$ largest singular values is
\begin{equation}
 \sum_{q=1}^{k}s_q(A_t)^2
 =t^2+m t^{-2}=2\sqrt m.
\end{equation}
Equations~\eqref{eq:rank-constrained-trace-bound} and
\eqref{eq:At-pairing} therefore give
\begin{equation}
 t\alpha+t^{-1}\beta
 \le \sqrt{2\sqrt m}\,\|C\|_{\mathrm F}.
\end{equation}
On the other hand,
\begin{equation}
 (t\alpha+t^{-1}\beta)^2\ge4\alpha\beta.
\end{equation}
Combining the last two inequalities yields
\begin{equation}
 2\alpha\beta\le\sqrt m\,\|C\|_{\mathrm F}^2,
\end{equation}
which is Eq.~\eqref{eq:collective-coherence-bound}.

To see that the constant is optimal, choose distinct
$j_1,\ldots,j_m\in J$ and take
\begin{equation}
 C=\frac{1}{\sqrt2} |1 \rangle \langle 1 |
 -\frac{1}{\sqrt{2m}}\sum_{q=1}^{m} |j_q \rangle \langle j_q|. 
\end{equation}
This matrix has rank $m+1\le k$, Frobenius norm one, and it saturates
Eq.~\eqref{eq:collective-coherence-bound}.
\end{proof}

\begin{proposition}[$k$-positivity]
\label{prop:kpos}
Let $2\le k\le d-1$. Then the map
$\Phi^{(a)}_{d,r}(w,z)$ is $k$-positive if and only if
\begin{equation}
\label{eq:kpos-criterion}
\begin{cases}
0\le w\le a, & r<d-1,\\[1mm]
0\le z\le \dfrac{a}{\sqrt{\min\{r,k-1\}}},
 & 1\le r\le d-1.
\end{cases}
\end{equation}
Consequently, when $r=1$, every level above positivity coincides with complete
positivity. At the opposite endpoint, $r=d-1$, the exchange sector is absent
and
\begin{equation}
 \widetilde W^{(a)}_d(z)\text{ is $k$-positive}
 \quad\Longleftrightarrow\quad
 0\le z\le\frac{a}{\sqrt{k-1}},
\end{equation}
so the intermediate $k$-positivity regions are genuinely distinct. The
remaining case $k=d$ is complete positivity and is covered by
Proposition~\ref{prop:cp}.
\end{proposition}

\begin{proof}
We use the Choi characterization of $k$-positivity recalled in
Sec.~\ref{sec:prelim}.

For necessity, suppose first that $r<d-1$, so that $I\ne\varnothing$. For any
$i\in I$, the Schmidt-rank-two vector
\begin{equation}
 \ket{\psi_i^-}:=\frac{\ket{1i}-\ket{i1}}{\sqrt2}
\end{equation}
satisfies
\begin{equation}
 \bra{\psi_i^-}W^{(a)}_{d,r}(w,z)\ket{\psi_i^-}=a-w.
\end{equation}
Thus $k$-positivity requires $w\le a$.

For the collective constraint, choose distinct indices
$j_1,\ldots,j_\ell\in J$ and define
\begin{equation}
 \ket{s_\ell}:=\frac1{\sqrt\ell}
 \sum_{q=1}^{\ell}\ket{j_qj_q}.
\end{equation}
On $\operatorname{span}\{\ket{11},\ket{s_\ell}\}$, the Choi operator is
represented by
\begin{equation}
 \begin{pmatrix}
  a&z\sqrt\ell\\
  z\sqrt\ell&a
 \end{pmatrix}.
\label{eq:kpositive-star-compression}
\end{equation}
Its smaller eigenvalue is $a-z\sqrt\ell$, with eigenvector proportional to
$\ket{11}-\ket{s_\ell}$. This vector has Schmidt rank $\ell+1$, so it is an
admissible Schmidt-rank-$k$ test whenever $\ell\le k-1$. Choosing
\begin{equation}
 \ell=\min\{r,k-1\}
\end{equation}
gives the necessary condition on $z$ in
Eq.~\eqref{eq:kpos-criterion}.

For sufficiency, let
\begin{equation}
 \ket\psi=\sum_{p,q=1}^{d}\psi_{pq}\ket{pq}
\end{equation}
have Schmidt rank at most $k$, and let $C_\psi=(\psi_{pq})$ be its coefficient
matrix. Then
\begin{equation}
 \rank C_\psi\le k,
 \qquad
 \|C_\psi\|_{\mathrm F}^2=\|\psi\|^2.
\end{equation}
Separate the matrix entries into the exchange pairs
$\{(1,i),(i,1)\}$, $i\in I$, the collective compression on the indices
$\{1\}\cup J$, and all remaining entries.

For each $i\in I$, the exchange contribution obeys
\begin{equation}
a\bigl(|\psi_{1i}|^2+|\psi_{i1}|^2\bigr)
+2w\Re(\overline{\psi_{1i}}\psi_{i1}) \ge
 (a-w)\bigl(|\psi_{1i}|^2+|\psi_{i1}|^2\bigr)\ge0, 
\label{eq:exchange-kpositive-bound}
\end{equation}
because $w \leq a$.
Let $C_J$ be the compression of $C_\psi$ to the rows and columns indexed by
$\{1\}\cup J$. Since compression cannot increase rank,
\begin{equation}
 \rank C_J\le k.
\end{equation}
With $m:=\min\{r,k-1\}$, Lemma~\ref{lem:collective-coherence} gives
\begin{equation}
 2\left|\overline{\psi_{11}}\sum_{j\in J}\psi_{jj}\right|
 \le\sqrt m\,\|C_J\|_{\mathrm F}^2.
\end{equation}
Therefore, if $z\le a/\sqrt m$,
\begin{equation}
a\|C_J\|_{\mathrm F}^2
 +2z\Re\!\left(
   \overline{\psi_{11}}\sum_{j\in J}\psi_{jj}
   \right)
     \ge
 (a-z\sqrt m)\|C_J\|_{\mathrm F}^2\ge0.
\label{eq:collective-kpositive-bound}
\end{equation}
All matrix entries not contained in the exchange pairs or in $C_J$ contribute
only the nonnegative diagonal term $a|\psi_{pq}|^2$. Adding these mutually
disjoint contributions yields
\begin{equation}
 \bra\psi W^{(a)}_{d,r}(w,z)\ket\psi\ge0.
\end{equation}
Hence the Choi operator is nonnegative on every vector of Schmidt rank at most
$k$, and the map is $k$-positive.
\end{proof}

\section{Decomposability and atomicity}
\label{sec:decomposability}
\label{sec:exactmixed}
The positivity hierarchy determines when the maps are valid, but it does not
identify which of their Choi witnesses can detect PPT entanglement. We now
resolve this question by deriving the exact decomposable and atomic regions. Throughout this section,
assume $1\le r\le d-2$ and retain the index sets $I,J$ and the
multiplicity $p$ defined in Sec.~\ref{sec:family}. Thus $p+r=d-1$, and
\begin{equation}
W^{(a)}_{d,r}(w,z)=aI_{d^2}
+w\sum_{i\in I}(\ket{1i}\!\bra{i1}+\ket{i1}\!\bra{1i})
\quad+z\sum_{j\in J}(\ket{11}\!\bra{jj}+\ket{jj}\!\bra{11}).
\label{eq:mixed-family}
\end{equation}
Define
\begin{equation}
\tau_w=\min\{w,a\},\quad\tau_z=\min\{z,a\},\quad
\mu_w=(w-a)_+,\quad\mu_z=(z-a)_+,
\label{eq:mixed-split}
\end{equation}
where $x_+:=\max\{x,0\}$ denotes the positive part. Clearly, $\tau_w+
\mu_w =w$ and $\tau_z +\mu_z =z$. 
\subsection{PPT obstruction to decomposability}
We first derive a necessary condition for decomposability by constructing PPT
operators adapted to the two channel classes and minimizing their expectation
value against the witness. For $x,y\ge0$, define
\begin{align}
R_{x,y}^{(d,r)}
&=\left(\ket{11}-x\sum_{j\in J}\ket{jj}\right)
\left(\bra{11}-x\sum_{j\in J}\bra{jj}\right)\nonumber\\
&\quad+x\sum_{j\in J}(\ket{1j}\!\bra{1j}+\ket{j1}\!\bra{j1})
+x^2\sum_{\substack{j,\ell\in J\\j<\ell}}
(\ket{j\ell}+\ket{\ell j})(\bra{j\ell}+\bra{\ell j})\nonumber\\
&\quad+y\sum_{i\in I}(\ket{1i}-\ket{i1})(\bra{1i}-\bra{i1})
+y^2\sum_{i\in I}\ket{ii}\!\bra{ii}\nonumber\\
&\quad+y^2\sum_{\substack{i,\ell\in I\\i<\ell}}
(\ket{i\ell}+\ket{\ell i})(\bra{i\ell}+\bra{\ell i}).
\label{eq:full-ppt}
\end{align}

\begin{lemma}[PPT property]\label{lem:full-ppt}
The operator $R_{x,y}^{(d,r)}$ is PPT for every $x,y\ge0$.
\end{lemma}
\begin{proof}
Positivity follows explicitly from \eqref{eq:full-ppt}. With
$\ket{\Omega_I}=\sum_{i\in I}\ket{ii}$ and
$\ket{\Omega_J}=\sum_{j\in J}\ket{jj}$, partial transposition gives
\begin{align}
(R_{x,y}^{(d,r)})^\Gamma
&=(\ket{11}-y\ket{\Omega_I})(\bra{11}-y\bra{\Omega_I})\nonumber\\
&\quad+y\sum_{i\in I}(\ket{1i}\!\bra{1i}+\ket{i1}\!\bra{i1})
+y^2\sum_{\substack{i,\ell\in I\\i<\ell}}
(\ket{i\ell}\!\bra{i\ell}+\ket{\ell i}\!\bra{\ell i})\nonumber\\
&\quad+x\sum_{j\in J}(\ket{1j}-\ket{j1})(\bra{1j}-\bra{j1})
+x^2\ket{\Omega_J}\!\bra{\Omega_J}\nonumber\\
&\quad+x^2\sum_{\substack{j,\ell\in J\\j<\ell}}
(\ket{j\ell}+\ket{\ell j})(\bra{j\ell}+\bra{\ell j}),
\label{eq:full-ppt-pt}
\end{align}
Every displayed term is positive semidefinite, so
$(R_{x,y}^{(d,r)})^\Gamma\ge0$.
\end{proof}

\begin{proposition}[Exact PPT threshold]\label{prop:full-detection}
One has
\begin{align}
\Tr [W^{(a)}_{d,r}(w,z)R_{x,y}^{(d,r)} ]
=a+ a r^2 x^2-2r(z-a)x+a p^2 y^2-2p(w-a)y.
\label{eq:full-pairing}
\end{align}
Consequently,
\begin{equation}
\inf_{x,y\ge0}\Tr [W^{(a)}_{d,r}(w,z)R_{x,y}^{(d,r)}]
=a-\frac{\mu_z^2+\mu_w^2}{a}.
\label{eq:full-inf}
\end{equation}
Hence the map $\Phi^{(a)}_{d,r}(w,z)$, equivalently its Choi witness
$W^{(a)}_{d,r}(w,z)$, is indecomposable if $\mu_z^2+\mu_w^2>a^2$.
\end{proposition}
\begin{proof}
The diagonal contribution from the collective sector is
\begin{equation}
 arx^2+2a\binom r2x^2=ar^2x^2,
\end{equation}
while the exchange sector contributes
\begin{equation}
 apy^2+2a\binom p2y^2=ap^2y^2.
\end{equation}
The terms linear in $x$ and $y$ are, respectively,
\begin{equation}
 2rax-2rzx=-2r(z-a)x,
 \qquad
 2pay-2pwy=-2p(w-a)y.
\end{equation}
Adding the contribution $a$ from $\ket{11}$ gives
Eq.~\eqref{eq:full-pairing}. Completing the squares,
\begin{align}
\Tr [W^{(a)}_{d,r}(w,z)R_{x,y}^{(d,r)}]
=a+ar^2\left(x-\frac{z-a}{ar}\right)^2
 +ap^2\left(y-\frac{w-a}{ap}\right)^2 
 -\frac{(z-a)^2+(w-a)^2}{a}.
\end{align}
Because $x,y\ge0$, the minimizing values are
\begin{equation}
 x_*=\frac{(z-a)_+}{ar}=\frac{\mu_z}{ar},
 \qquad
 y_*=\frac{(w-a)_+}{ap}=\frac{\mu_w}{ap},
\end{equation}
which yields Eq.~\eqref{eq:full-inf}.
\end{proof}

\subsection{Explicit decomposable representation}
The PPT family proves indecomposability outside a boundary, namely
$\mu_z ^2 + \mu_w ^2 >a^2$. To establish sufficiency on the
complementary side, we now construct an explicit representation of the
form in Eq.~\eqref{eq:decomposable-form}.

\begin{proposition}[Collective decomposition]\label{prop:collective-decomp}
If $\mu_z^2+\mu_w^2\le a^2$, then the map
$\Phi^{(a)}_{d,r}(w,z)$, equivalently $W^{(a)}_{d,r}(w,z)$, is
decomposable.
\end{proposition}
\begin{proof}
Put
\begin{equation}
\lambda=a-\frac{\mu_z^2+\mu_w^2}{a},\quad
\ket{u_z}=\frac{\mu_z}{\sqrt a}\ket{11}
+\sqrt a\sum_{j\in J}\ket{jj},\quad
\ket{v_w}=\frac{\mu_w}{\sqrt a}\ket{11}
+\sqrt a\sum_{i\in I}\ket{ii}.
\end{equation}
Define
\begin{align}
P&=\ket{u_z}\!\bra{u_z}+\lambda\ket{11}\!\bra{11}
+\tau_w\sum_{i\in I}(\ket{1i}+\ket{i1})(\bra{1i}+\bra{i1})\nonumber\\
&\quad+a\sum_{\substack{i,\ell\in I\\i<\ell}}
(\ket{i\ell}-\ket{\ell i})(\bra{i\ell}-\bra{\ell i})\nonumber\\
&\quad+(a-\tau_z)\sum_{j\in J}
(\ket{1j}\!\bra{1j}+\ket{j1}\!\bra{j1})
+a\sum_{\substack{i\in I\\j\in J}}
(\ket{ij}\!\bra{ij}+\ket{ji}\!\bra{ji}),
\label{eq:P-all}
\end{align}
and
\begin{align}
Q&=\ket{v_w}\!\bra{v_w}
+(a-\tau_w)\sum_{i\in I}
(\ket{1i}\!\bra{1i}+\ket{i1}\!\bra{i1})\nonumber\\
&\quad+\tau_z\sum_{j\in J}(\ket{1j}+\ket{j1})(\bra{1j}+\bra{j1})\nonumber\\
&\quad+a\sum_{\substack{j,\ell\in J\\j<\ell}}
(\ket{j\ell}-\ket{\ell j})(\bra{j\ell}-\bra{\ell j}).
\label{eq:Q-all}
\end{align}
Clearly $P,Q\ge0$. To verify the decomposition, first note that
\begin{align}
\ket{u_z}\!\bra{u_z}
&=\frac{\mu_z^2}{a}\ket{11}\!\bra{11}
+a\sum_{j,\ell\in J}\ket{jj}\!\bra{\ell\ell}
+\mu_z\sum_{j\in J}
 (\ket{11}\!\bra{jj}+\ket{jj}\!\bra{11}),\label{eq:ualpha-expand}\\
(\ket{v_w}\!\bra{v_w})^\Gamma
&=\frac{\mu_w^2}{a}\ket{11}\!\bra{11}
+a\sum_{i,\ell\in I}\ket{i\ell}\!\bra{\ell i}
+\mu_w\sum_{i\in I}
 (\ket{1i}\!\bra{i1}+\ket{i1}\!\bra{1i}).
\label{eq:vdelta-expand}
\end{align}
For $i<\ell$ in $I$, the off-diagonal terms
$a(\ket{i\ell}\!\bra{\ell i}+\ket{\ell i}\!\bra{i\ell})$ from
Eq.~\eqref{eq:vdelta-expand} are canceled by the corresponding terms in
\begin{equation}
 a(\ket{i\ell}-\ket{\ell i})(\bra{i\ell}-\bra{\ell i})
\end{equation}
of $P$. Likewise, for $j<\ell$ in $J$, the off-diagonal terms
$a(\ket{jj}\!\bra{\ell\ell}+\ket{\ell\ell}\!\bra{jj})$ from
Eq.~\eqref{eq:ualpha-expand} are canceled by the partial transpose of the
antisymmetric $J$-sector terms in $Q$. The remaining $I$--$J$ diagonal entries
are supplied by the last sum in $P$. Finally,
\begin{equation}
 \frac{\mu_z^2}{a}+\lambda+\frac{\mu_w^2}{a}=a,
 \qquad
 \tau_w+\mu_w=w,
 \qquad
 \tau_z+\mu_z=z.
\end{equation}
These identities account for every nonzero matrix element and prove
the decomposition
\begin{equation}
 W^{(a)}_{d,r}(w,z)=P+Q^\Gamma.
\end{equation}
\end{proof}

\begin{proposition}[Exact decomposability boundary]\label{thm:exact-all}
Let $d\ge3$, $1\le r\le d-2$, and $0\le w,z\le2a$. Then
\begin{equation}
W^{(a)}_{d,r}(w,z)\text{ is decomposable}
\Longleftrightarrow
(w-a)_+^2+(z-a)_+^2\le a^2.
\label{eq:exact-all}
\end{equation}
\end{proposition}
\begin{proof}
Proposition~\ref{prop:collective-decomp} proves sufficiency, and
Proposition~\ref{prop:full-detection} proves necessity.
\end{proof}

The absence of $p$ and $r$ from Eq.~\eqref{eq:exact-all} is nontrivial:
the two multiplicities enter the PPT trial family and the decomposition
separately, but their collective contributions balance at the optimum.

\subsection{Atomicity of the mixed families}
\label{sec:atomicity}
The preceding proposition locates the indecomposable region, but
indecomposability does not imply atomicity in general. For the present mixed
families, a three-dimensional subspace retaining one channel of each type
provides a Schmidt-number-two PPT certificate throughout that region. Choose $i_0\in I$ and $j_0\in J$. For $s,c\ge0$, $s^2+c^2=1$, define
\begin{align}
\mathcal A_{s,c}&=(\ket{11}-s\ket{j_0j_0})(\bra{11}-s\bra{j_0j_0})\nonumber\\
&\quad+c(\ket{1i_0}-\ket{i_01})(\bra{1i_0}-\bra{i_01})
+c^2\ket{i_0i_0}\!\bra{i_0i_0}\nonumber\\
&\quad+s\ket{1j_0}\!\bra{1j_0}+s\ket{j_01}\!\bra{j_01}.
\end{align}
Its partial transpose can be written explicitly as
\begin{align}
\mathcal A_{s,c}^\Gamma
&=(\ket{11}-c\ket{i_0i_0})(\bra{11}-c\bra{i_0i_0})\nonumber\\
&\quad+c(\ket{1i_0}\!\bra{1i_0}+\ket{i_01}\!\bra{i_01})\nonumber\\
&\quad+s(\ket{1j_0}-\ket{j_01})(\bra{1j_0}-\bra{j_01})
+s^2\ket{j_0j_0}\!\bra{j_0j_0}.
\label{eq:embedded-qutrit-pt}
\end{align}
Both $\mathcal A_{s,c}$ and $\mathcal A_{s,c}^\Gamma$ are positive and have Schmidt
number at most two. Since $ \Tr \mathcal A_{s,c}=2(1+s+c)$, the normalized PPT
state
\begin{equation}
 \rho_{s,c}:=\frac{\mathcal A_{s,c}}{2(1+s+c)}\label{rsc}
\end{equation}
satisfies
\begin{equation}
\Tr\!\left[W^{(a)}_{d,r}(w,z)\rho_{s,c}\right]
=\frac{a(1+s+c)-zs-wc}{1+s+c}.
\label{eq:embedded-qutrit-expectation}
\end{equation} 
If $\mu_z^2+\mu_w^2>a^2$, choose
$s=\mu_z/\sqrt{\mu_z^2+\mu_w^2}$ and
$c=\mu_w/\sqrt{\mu_z^2+\mu_w^2}$. The expectation is negative.

\begin{proposition}[Exact atomicity boundary]\label{thm:atomic-all}
For every block-positive mixed member of the family with $1\le r\le d-2$,
\begin{equation}
W^{(a)}_{d,r}(w,z)\text{ is atomic}
\Longleftrightarrow W^{(a)}_{d,r}(w,z)\text{ is indecomposable}
\Longleftrightarrow
(w-a)_+^2+(z-a)_+^2>a^2.
\end{equation}
\end{proposition}
\begin{proof}
The embedded state and its partial transpose have Schmidt number at most two,
so negative expectation implies atomicity. Conversely, every atomic map is
indecomposable, and Proposition~\ref{thm:exact-all} gives the exact region.
\end{proof}

\subsection{The purely collective endpoint}
\label{sec:star}
The mixed-family proof relies on the coexistence of exchange and collective
coherences, weighted by $w$ and $z$, respectively. When the exchange sector disappears, the behavior changes
qualitatively.

\begin{proposition}[Purely collective endpoint]
\label{prop:star-endpoint}
For $r=d-1$, the operator $\widetilde W^{(a)}_d(z)$ in
Eq.~\eqref{eq:tilde-family} is block positive and decomposable if and only if
$0\le z\le2a$. Consequently, this endpoint is never atomic in its
block-positive region.
\end{proposition}
\begin{proof}
Block positivity is Proposition~\ref{prop:positivity}. To prove
decomposability, put
\begin{equation}
 \tau_z:=\min\{z,a\},\qquad
 \mu_z:=(z-a)_+,
 \qquad
 \lambda:=a-\frac{\mu_z^2}{a}.
\end{equation}
For $0\le z\le2a$, one has $0\le\mu_z\le a$ and hence $\lambda\ge0$. Define
\begin{align}
P_\star&=
 \left(\frac{\mu_z}{\sqrt a}\ket{11}
 +\sqrt a\sum_{j=2}^d\ket{jj}\right)
 \left(\frac{\mu_z}{\sqrt a}\bra{11}
 +\sqrt a\sum_{j=2}^d\bra{jj}\right)
 +\lambda\ket{11}\!\bra{11}\nonumber\\
&\quad +(a-\tau_z)\sum_{j=2}^d
 (\ket{1j}\!\bra{1j}+\ket{j1}\!\bra{j1}),\\
Q_\star&=
 \tau_z\sum_{j=2}^d(\ket{1j}+\ket{j1})(\bra{1j}+\bra{j1})\nonumber\\
&\quad+a\sum_{2\le j<\ell\le d}
 (\ket{j\ell}-\ket{\ell j})(\bra{j\ell}-\bra{\ell j}).
\end{align}
Both operators are positive semidefinite. Moreover,
\begin{align}
Q_\star^\Gamma
&=\tau_z\sum_{j=2}^d
 (\ket{1j}\!\bra{1j}+\ket{j1}\!\bra{j1}
 +\ket{11}\!\bra{jj}+\ket{jj}\!\bra{11})\nonumber\\
&\quad+a\sum_{2\le j<\ell\le d}
 (\ket{j\ell}\!\bra{j\ell}+\ket{\ell j}\!\bra{\ell j}
 -\ket{jj}\!\bra{\ell\ell}-\ket{\ell\ell}\!\bra{jj}).
\label{eq:Qstar-pt}
\end{align}
The off-diagonal coherences among the vectors $\ket{jj}$ in $P_\star$ are canceled by the last two terms in
Eq.~\eqref{eq:Qstar-pt}. The remaining entries reproduce the diagonal part
and the collective coherences. Thus
\begin{equation}
 \widetilde W^{(a)}_d(z)=P_\star+Q_\star^\Gamma,
\end{equation}
which proves decomposability.
\end{proof}

\subsection{Complete phase diagram}
\label{sec:summary}
\begin{figure}[t]
 \centering
 \includegraphics[width=0.9\textwidth]{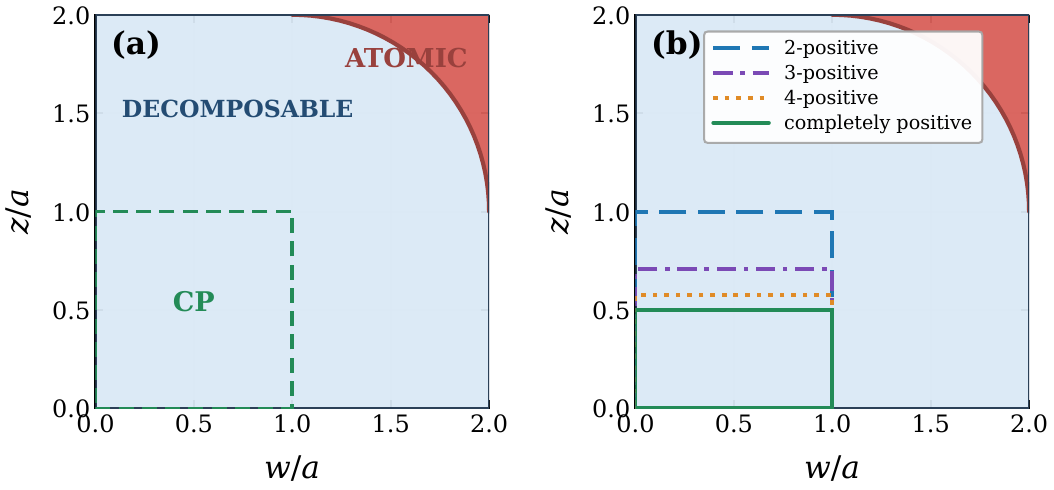}
 \caption{Exact phase structure of the mixed families in the scale-free
coordinates $(w/a,z/a)$. The blue square is the block-positive region, the red
cap is atomic and indecomposable, and the intervening region is decomposable;
the red arc is the exact decomposability boundary. (a) For $r=1$, the dashed
green square is the completely positive region. (b) For $r=4$,
the nested rectangles are the $k$-positive regions, and the smallest
rectangle corresponds to complete positivity. Thus $r$ changes the $k$-positivity
hierarchy but not the decomposability boundary.}
 \label{fig:phase-diagram-all-r}
\end{figure}

We now assemble the preceding classifications into a single phase diagram. For
$1\le r\le d-2$, the exact conditions are
\begin{align}
W^{(a)}_{d,r}(w,z)\text{ block positive}
&\Longleftrightarrow 0\le w,z\le2a,\\
W^{(a)}_{d,r}(w,z)\text{ completely positive}
&\Longleftrightarrow w\le a,\quad z\le\frac{a}{\sqrt r},\\
W^{(a)}_{d,r}(w,z)\text{ $k$-positive}
&\Longleftrightarrow w\le a,\quad
z\le\frac{a}{\sqrt{\min\{r,k-1\}}},\\
W^{(a)}_{d,r}(w,z)\text{ decomposable}
&\Longleftrightarrow (w-a)_+^2+(z-a)_+^2\le a^2,\\
W^{(a)}_{d,r}(w,z)\text{ atomic}
&\Longleftrightarrow (w-a)_+^2+(z-a)_+^2>a^2.
\end{align}
Hence $r$ changes complete and higher positivity, but not the exact
decomposability or atomicity boundary of a mixed family. 
Figure~\ref{fig:phase-diagram-all-r} summarizes these results in the scale-free
coordinates $w/a$ and $z/a$. Panel~(a) displays the universal positivity square,
the exact decomposable region, and the atomic cap. Panel~(b) illustrates the
$r$-dependent nesting of the $k$-positivity regions for the representative
case $r=4$; the exact circular decomposability boundary is unchanged.

\section{PPT edge states and detection geometry}
\label{sec:edge-states}
The PPT operators introduced in Sec.~\ref{sec:decomposability} to locate the
failure of decomposability also carry
useful state-space information. After normalization, they provide an
explicit family of PPT entangled states whose boundary character and detection
geometry can be determined analytically.

\subsection{Edge property}
Consider the PPT operators introduced in Eq.~\eqref{eq:full-ppt}. Put
\begin{equation}
 \rho_{x,y}^{(d,r)}
 :=\frac{R_{x,y}^{(d,r)}}{N_{x,y}^{(d,r)}},
 \qquad
 N_{x,y}^{(d,r)}:=\Tr R_{x,y}^{(d,r)}.
\end{equation}
A direct count of the rank-one summands gives
\begin{equation}
 N_{x,y}^{(d,r)}
 =1+2rx+r^2x^2+2py+p^2y^2.
\label{eq:edge-normalization}
\end{equation}
Normalization does not change either range and therefore does not affect the
edge property.

\begin{proposition}[Edge property]
\label{prop:full-edge}
Let $d\ge3$, $1\le r\le d-2$, and $x,y>0$. Then
$\rho_{x,y}^{(d,r)}$ is a PPT edge state.
\end{proposition}
\begin{proof}
The PPT property was proved in Lemma~\ref{lem:full-ppt}. Suppose that a
product vector $\ket f\otimes\ket g$ satisfies
\begin{equation}
 \ket f\otimes\ket g\in\Ran R_{x,y}^{(d,r)},
 \qquad
 \ket f\otimes\ket{\overline g}
 \in\Ran (R_{x,y}^{(d,r)})^\Gamma.
\label{eq:compatible-product}
\end{equation}
Write the components of $f$ and $g$ relative to the distinguished index $1$
and the sets $I,J$. Orthogonality to the kernels, or equivalently direct use
of the range decompositions in Eqs.~\eqref{eq:full-ppt} and
\eqref{eq:full-ppt-pt} give
\begin{align}
 f_i g_j&=f_jg_i=0,
 &&i\in I,\ j\in J,\label{eq:edge-mixed}\\
 f_1g_i+f_i g_1&=0,
 &&i\in I,\label{eq:edge-I1}\\
 f_i g_\ell&=f_\ell g_i,
 &&i,\ell\in I,\label{eq:edge-II}\\
 f_jg_j&=-x f_1g_1,
 &&j\in J,\label{eq:edge-Jdiag}\\
 f_1\overline{g_j}+f_j\overline{g_1}&=0,
 &&j\in J,\label{eq:edge-J1}\\
 f_i\overline{g_i}&=-y f_1\overline{g_1},
 &&i\in I,\label{eq:edge-Idiag}\\
 f_j\overline{g_\ell}&=f_\ell\overline{g_j},
 &&j,\ell\in J,\label{eq:edge-JJ}\\
 f_j\overline{g_j}&=f_\ell\overline{g_\ell},
 &&j,\ell\in J.\label{eq:edge-Jdiag-pt}
\end{align}
The last condition records the $\ket{\Omega_J}$ component of
$\Ran(R_{x,y}^{(d,r)})^\Gamma$: the diagonal products
$f_j\overline{g_j}$ must be independent of $j\in J$. The contradiction below
uses only necessary range conditions, so this additional constraint is not
needed in the subsequent case analysis.
If $f_1g_1\ne0$, Eqs.~\eqref{eq:edge-Jdiag} and
\eqref{eq:edge-Idiag}, together with $x,y>0$, force nonzero components in both
$I$ and $J$, contradicting Eq.~\eqref{eq:edge-mixed}. Hence $f_1g_1=0$.
If exactly one of $f_1,g_1$ vanishes, Eqs.~\eqref{eq:edge-I1} and
\eqref{eq:edge-J1} force the factor whose first component vanishes to be the
zero vector. For example, if $f_1=0$ and $g_1\ne0$, those equations give
$f_i=f_j=0$ for every $i\in I$ and $j\in J$, hence $f=0$; the case
$g_1=0$ and $f_1\ne0$ is symmetric. Thus
$f_1=g_1=0$.

We now use Eq.~\eqref{eq:edge-mixed} to determine the remaining support.
If some $f_i\neq0$ with $i\in I$, then $g_j=0$ for every $j\in J$;
if some $g_i\neq0$ with $i\in I$, then $f_j=0$ for every $j\in J$.
The analogous statements follow after interchanging $I$ and $J$. Hence a
nonzero compatible product vector must be supported entirely in
$I\otimes I$ or entirely in $J\otimes J$.

Suppose first that the support lies in $I\otimes I$. Equation~\eqref{eq:edge-II}
means that all $2\times2$ minors of the two-row array
$(f_I;g_I)$ vanish. Thus, unless one restricted vector is zero, there is an
$\zeta\in\mathbb C$ such that
\begin{equation}
 f_i=\zeta g_i\qquad(i\in I).
\end{equation}
Because $f_1=g_1=0$, Eq.~\eqref{eq:edge-Idiag} gives
\begin{equation}
 0=f_i\overline{g_i}=\zeta|g_i|^2\qquad(i\in I),
\end{equation}
which forces $f_I=0$ or $g_I=0$. In either case the product vector vanishes.

If the support lies in $J\otimes J$, Eq.~\eqref{eq:edge-JJ} similarly implies
$f_j=\zeta\overline{g_j}$ for all $j\in J$. Equation~\eqref{eq:edge-Jdiag}
then yields
\begin{equation}
 0=f_jg_j=\zeta|g_j|^2\qquad(j\in J),
\end{equation}
and again one factor must vanish. Both alternatives contradict the assumption
that the product vector is nonzero. Therefore no vector satisfying
Eq.~\eqref{eq:compatible-product} exists, and the PPT state is an edge
state.
\end{proof}

\subsection{Detection region of a fixed witness}
Having established the edge property, we next determine which members of
the family in Eq.~\eqref{eq:full-ppt} are detected by a fixed block-positive
witness $W^{(a)}_{d,r}(w,z)$. Normalization is positive, so
$\rho_{x,y}^{(d,r)}$ is detected exactly when the unnormalized
expectation in Eq.~\eqref{eq:full-pairing} is negative. Completing
the squares gives the following result.

\begin{proposition}[Detected edge-state region]
\label{prop:detected-edge-region}
Let $p=d-r-1$ and define
\begin{equation}
 x_0:=\frac{z-a}{ra},
 \qquad
 y_0:=\frac{w-a}{pa}.
\label{eq:ellipse-center}
\end{equation}
The witness $W^{(a)}_{d,r}(w,z)$ detects the states with $x,y>0$ satisfying
\begin{equation}
 r^2a(x-x_0)^2+p^2a(y-y_0)^2
 <\frac{(z-a)^2+(w-a)^2-a^2}{a}.
\label{eq:detected-ellipse}
\end{equation}
Thus the detected region is the intersection of the open ellipse with the
open positive quadrant $x,y>0$. This intersection is nonempty if and only if
\begin{equation}
 (z-a)_+^2+(w-a)_+^2>a^2,
\end{equation}
i.e., when the map is atomic and indecomposable.
\end{proposition}
\begin{proof}
Starting from Eq.~\eqref{eq:full-pairing}, complete the squares separately in
$x$ and $y$. This gives Eq.~\eqref{eq:detected-ellipse}. Minimization over
the positive quadrant replaces $z-a$ and $w-a$ by their positive parts, and
the minimum is the value in Eq.~\eqref{eq:full-inf}. Hence the ellipse has a
nonempty intersection with $x,y>0$ exactly in the stated region.
\end{proof}

The geometry simplifies in the collective coordinates
\begin{equation}
 X:=rx,
 \qquad
 Y:=py.
\end{equation}
Equation~\eqref{eq:detected-ellipse} becomes the disk
\begin{equation}
 \bigl(X-(z/a-1)\bigr)^2
 +\bigl(Y-(w/a-1)\bigr)^2
 <(z/a-1)^2+(w/a-1)^2-1,
 \qquad X,Y>0.
\label{eq:detected-disk}
\end{equation}
Therefore the multiplicities $p$ and $r$ change the semiaxes in the original
$(x,y)$ plane but disappear after the natural collective rescaling. On the
exact decomposability boundary, the intersection of the ellipse with the open
quadrant is empty. If $w,z\ge a$, the ellipse degenerates to the single
zero-expectation point
\begin{equation}
 x_*=\frac{z-a}{ra},
 \qquad
 y_*=\frac{w-a}{pa}.
\end{equation}
If one coupling is smaller than $a$, the boundary ellipse need not collapse;
instead, it is tangent to the corresponding coordinate axis and has no point
with $x,y>0$. Inside the decomposable region the detected set is empty as
well.

\begin{figure}[t]
 \centering
 \includegraphics[width=0.9\textwidth]{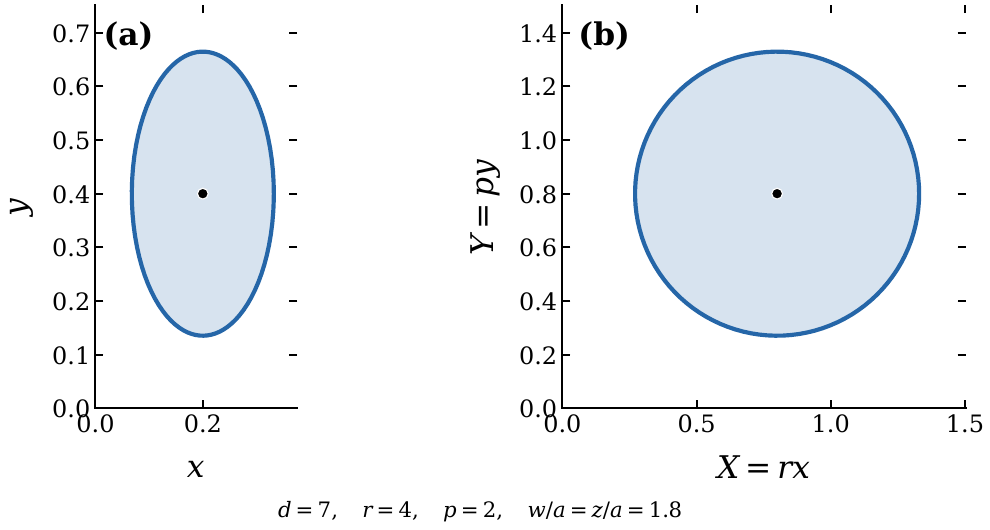}
 \caption{Detection geometry of the PPT edge states for
$W^{(a)}_{7,4}(w,z)$ at $w/a=z/a=1.8$, with multiplicities $p=2$ and $r=4$.
(a) In the state parameters $(x,y)$, the states with negative witness
expectation lie inside the ellipse. (b) In the collective coordinates $X=rx$
and $Y=py$, the same boundary is a circle. In both panels, the solid curve is
the zero-expectation contour and the dot marks its center.}
 \label{fig:detected-edge-region}
\end{figure}

Figure~\ref{fig:detected-edge-region} illustrates both descriptions for the
representative choice $d=7$, $r=4$, and $w/a=z/a=1.8$.

\section{Optimal refinement at maximal coherence}
\label{sec:optimal-refinement}
The phase diagram determines atomicity, but atomicity does not imply
optimality. We complete the analysis by refining the maximally coherent mixed witness
through an explicit positive subtraction.

\subsection{Construction of the refined witness}
The original witnesses do not generally have the spanning property. At the
maximally coherent mixed endpoint, however, a natural positive subtraction
produces a witness with the spanning property, which is then optimal. Throughout this section, let
$1\le r\le d-2$, retain the sets $I$ and $J$ defined in
Sec.~\ref{sec:family}, and denote by
\begin{equation}
 W^{\max,(a)}_{d,r}:=W^{(a)}_{d,r}(2a,2a)
\end{equation}
the maximally coherent witness. Define the positive operator
\begin{equation}
D^{(a)}_{d,r}
:=a\sum_{i\in I}
 (\ket{1i}+\ket{i1})(\bra{1i}+\bra{i1})\quad
+a\sum_{\substack{i,\ell\in I\\i<\ell}}
 (\ket{i\ell}-\ket{\ell i})(\bra{i\ell}-\bra{\ell i}),
\label{eq:refinement-subtraction}
\end{equation}
and the refined witness
\begin{equation}
 \widehat W^{(a)}_{d,r}:=W^{\max,(a)}_{d,r}-D^{(a)}_{d,r}.
\label{eq:refined-witness}
\end{equation}
For $d=3$ and $r=1$, the second sum in
Eq.~\eqref{eq:refinement-subtraction} is empty, and the construction reduces
to the optimal refinement of the qutrit endpoint.

\begin{proposition}[Optimal refinement]
\label{thm:optimal-refinement}
For every $d\ge3$ and $1\le r\le d-2$, the operator
$\widehat W^{(a)}_{d,r}$ is block positive and its zero product vectors span
$\mathbb C^d\otimes\mathbb C^d$. Consequently,
$\widehat W^{(a)}_{d,r}$ is an optimal entanglement witness. Moreover, it remains
indecomposable and atomic.
\end{proposition}

\subsection{Block positivity}
We first show that the subtraction preserves block positivity by reducing the
product expectation to a positive compression.

\begin{proof}[Proof of block positivity]
Fix a unit vector $\ket{x}$ and decompose it according to
\begin{equation}
 \mathbb C^d
 =\operatorname{span}\{\ket1\}\oplus\mathcal H_I\oplus\mathcal H_J,
\end{equation}
where
\begin{equation}
 \mathcal H_I:=\operatorname{span}\{\ket i:i\in I\},
 \qquad
 \mathcal H_J:=\operatorname{span}\{\ket j:j\in J\}.
\end{equation}
Write
\begin{equation}
 \ket{x}=\xi \ket1+\ket u+\ket v,
 \qquad
 u\in\mathcal H_I,
 \quad
 v\in\mathcal H_J,
\end{equation}
and set
\begin{equation}
 A:=|\xi|^2,
 \qquad
 U:=\|u\|^2,
 \qquad
 V:=\|v\|^2.
\end{equation}
Since $x$ is normalized,
\begin{equation}
 A+U+V=1.
\label{eq:refinement-normalization}
\end{equation}
Block positivity of $\widehat W^{(a)}_{d,r}$ is equivalent to positivity of the
operator
\begin{equation}
 B_x:=\bra{x}\widehat W^{(a)}_{d,r}\ket{x}\label{bx}
\end{equation}
on the second tensor factor. For an arbitrary vector
$\ket y=\gamma\ket1+\ket{s}+\ket{t}$, with
$s\in\mathcal H_I$ and $t\in\mathcal H_J$, the two contributions can be
written separately. From the maximally coherent witness,
\begin{align}
\frac1a\bra{x,y}W^{\max,(a)}_{d,r}\ket{x,y}
&=\|y\|^2
 +4\Re\!\left(\overline\gamma\,\xi\langle u|s\rangle
 +\overline\gamma\,\overline\xi\langle\overline v|t\rangle\right),
\label{eq:Wmax-quadratic-form}
\end{align}
whereas the subtraction gives
\begin{equation}
\frac1a\bra{x,y}D^{(a)}_{d,r}\ket{x,y}
=U|\gamma|^2+(A+U)\|s\|^2-|\langle u|s\rangle|^2 +2\Re\!\left(\overline\gamma\,\xi\langle u|s\rangle\right).
\label{eq:D-quadratic-form}
\end{equation}
Here the identity
\begin{equation}
 \sum_{i<\ell}|u_i s_\ell-u_\ell s_i|^2
 =U\|s\|^2-|\langle u|s\rangle|^2
\end{equation}
has been used for the antisymmetric $I$--$I$ sector. Subtracting
Eq.~\eqref{eq:D-quadratic-form} from Eq.~\eqref{eq:Wmax-quadratic-form}, and
using $A+U+V=1$, yields
\begin{equation}
\frac1a\bra yB_x\ket y
=(1-U)|\gamma|^2+V\|s\|^2+\|t\|^2+|\langle u|s\rangle|^2
+2\Re\!\left(\overline\gamma\,\xi\langle u|s\rangle
 +2\overline\gamma\,\overline\xi\langle\overline v|t\rangle\right).
\label{eq:Bx-quadratic-form}
\end{equation}
Therefore, in the decomposition
$\operatorname{span}\{\ket1\}\oplus\mathcal H_I\oplus\mathcal H_J$,
the operator $B_x/a$ is represented by
\begin{equation}
 \frac1aB_x=
 \begin{pmatrix}
 1-U& \xi u^\dagger&2\overline \xi\,v^T\\
 \overline \xi\,u&VI_p+uu^\dagger&0\\
 2\xi\,\overline v&0&I_r
 \end{pmatrix}.
\label{eq:refinement-compression}
\end{equation}
The lower-right block is $I_r>0$. By the Schur-complement criterion
\cite{HornJohnson2013}, positivity is therefore equivalent to positivity of
\begin{equation}
 M_x=
 \begin{pmatrix}
 1-U-4 A V&\xi u^\dagger\\
 \overline \xi\,u&VI_p+uu^\dagger
 \end{pmatrix}.
\label{eq:first-schur-refinement}
\end{equation}
Suppose first that $V>0$. Then
\begin{equation}
 G:=VI_p+uu^\dagger>0,
\end{equation}
and the Sherman--Morrison formula
\cite{ShermanMorrison1950,HornJohnson2013} gives
\begin{equation}
 G^{-1}
 =\frac1V\left(I_p-\frac{uu^\dagger}{U+V}\right),
 \qquad
 u^\dagger G^{-1}u=\frac{U}{U+V}.
\end{equation}
The remaining scalar Schur complement is therefore
\begin{equation}
 S_x
 =1-U-4 A V-X\frac{U}{U+V}
 =\frac{V(2A-1)^2}{1-A}\ge0,
\label{eq:final-schur-refinement}
\end{equation}
where Eq.~\eqref{eq:refinement-normalization} was used. 
Thus $B_x\ge0$ whenever $V>0$.

If $V=0$, then $A+U=1$ and Eq.~\eqref{eq:first-schur-refinement} becomes
\begin{equation}
 M_x=
 \begin{pmatrix}
 A&\xi u^\dagger\\
 \overline \xi\,u&uu^\dagger
 \end{pmatrix}
 =
 \begin{pmatrix}\xi\\u\end{pmatrix}
 \begin{pmatrix}\xi\\u\end{pmatrix}^{\!\dagger}\ge0.
\end{equation}
Hence $B_x\ge0$ for every $x$, which
proves
\begin{equation}
 \bra{x,y}\widehat W^{(a)}_{d,r}\ket{x,y}\ge0
 \qquad\text{for all }x,y.
\end{equation}
Therefore $\widehat W^{(a)}_{d,r}$ is block positive.
\end{proof}

\subsection{Spanning property and optimality}
Having established that the subtraction is admissible, we prove optimality by
verifying the spanning property defined in Sec.~\ref{sec:prelim}. The
Schur-complement argument identifies a spanning set of zero product vectors.
Put
\begin{equation}
 \mathcal K:=\operatorname{span}\{\ket1\}\oplus\mathcal H_I.
\end{equation}
Since $I$ is nonempty, $\dim\mathcal K=p+1\ge2$.

\begin{lemma}
\label{lem:orthogonal-products-span}
If $\dim\mathcal K\ge2$, then
\begin{equation}
 \operatorname{span}\{\ket{x}\otimes\ket{y}:\ket{x},\ket{y}\in\mathcal K,
 \ \langle x|y\rangle=0\}
 =\mathcal K\otimes\mathcal K.
\label{eq:orthogonal-products-span}
\end{equation}
\end{lemma}
\begin{proof}
Let $\{\ket{e_\mu}\}$ be an orthonormal basis of $\mathcal K$. For
$\mu\ne\nu$, the tensor $\ket{e_\mu}\otimes\ket{e_\nu}$ already belongs to
the set on the left-hand side. The orthogonal pairs
$\ket{e_\mu+e_\nu},\ket{e_\mu-e_\nu}$ and
$\ket{e_\mu+i e_\nu},\ket{i e_\mu+e_\nu}$ give the product tensors
\begin{align}
&(\ket{e_\mu}+\ket{e_\nu})\otimes
 (\ket{e_\mu}-\ket{e_\nu}),\nonumber\\
&(\ket{e_\mu}+i\ket{e_\nu})\otimes
 (i\ket{e_\mu}+\ket{e_\nu}).
\end{align}
Subtracting the off-diagonal tensors, which are already in the span, leaves
\begin{equation}
 \ket{e_\mu e_\mu}-\ket{e_\nu e_\nu},
 \qquad
 i\bigl(\ket{e_\mu e_\mu}+\ket{e_\nu e_\nu}\bigr).
\end{equation}
Their sum and difference isolate both diagonal tensors. Hence every basis
tensor $\ket{e_\mu}\otimes\ket{e_\nu}$ lies in the span.
\end{proof}

\begin{proof}[Proof of the spanning property]
If $v=0$, then $V=0$ and Eq.~\eqref{eq:refinement-compression} gives, on
$\mathcal K=\operatorname{span}\{\ket1\}\oplus\mathcal H_I$,
\begin{equation}
 \frac1a B_x\big|_{\mathcal K}
 =\begin{pmatrix}
 |\xi|^2&\xi u^\dagger\\
 \overline\xi\,u&uu^\dagger
 \end{pmatrix}
 =\begin{pmatrix}\xi\\u\end{pmatrix}
 \begin{pmatrix}\xi\\u\end{pmatrix}^{\!\dagger}
 =\ket{x}\!\bra{x}.
\label{eq:Bx-vzero}
\end{equation}
Consequently,
\begin{equation}
 \bra{x,y}\widehat W^{(a)}_{d,r}\ket{x,y}=0
 \qquad
 \text{whenever }x,y\in\mathcal K,
 \quad\langle x|y\rangle=0.
\end{equation}
Lemma~\ref{lem:orthogonal-products-span} therefore implies that the zero
product vectors span $\mathcal K\otimes\mathcal K$.

A second family of zero product vectors follows from the equality case of
Eq.~\eqref{eq:final-schur-refinement}. Choose
\begin{equation}
 |\xi |^2=\frac12,
 \qquad
 \|u\|^2+\|v\|^2=\frac12,
 \qquad
 V>0.
\label{eq:second-zero-constraints}
\end{equation}
Then the vector
\begin{equation}
 \ket y
 =\ket1-2\overline\xi\,\ket u-2\xi\ket{\overline v}
\label{eq:second-zero-vector}
\end{equation}
belongs to the kernel of $B_x$, where
\begin{equation}
 \ket x=\xi\ket1+\ket u+\ket v.
\end{equation}
Hence $\ket{x}\otimes\ket{y}$ is a zero product vector.

Write $\xi=e^{i\theta}/\sqrt2$. Since every vector in
$\mathcal K\otimes\mathcal K$ is already in the span of zero product
vectors, we may discard from $\ket{x}\otimes\ket{y}$ all terms belonging to
that subspace. We write $\equiv$ for equality modulo
$\mathcal K\otimes\mathcal K$. The remaining component is
\begin{equation}
 \ket{Z(\theta;u,v)}
 \equiv \ket v\otimes\ket1
 -2\xi^2\ket1\otimes\ket{\overline v}
 -2\xi\,\ket u\otimes\ket{\overline v}
 -2\overline\xi\,\ket v\otimes\ket u
 -2\xi\,\ket v\otimes\ket{\overline v}.
\label{eq:fourier-zero-vector}
\end{equation}
The terms in Eq.~\eqref{eq:fourier-zero-vector} have distinct Fourier modes
in $\theta$. Projecting onto the distinct Fourier coefficients of $e^{im\theta}$ isolates
\begin{equation}
 \ket v\otimes\ket1,
 \qquad
 \ket1\otimes\ket{\overline v},
 \qquad
 \ket v\otimes\ket u,
 \qquad
 \ket u\otimes\ket{\overline v}+\ket v\otimes\ket{\overline v}.
\label{eq:isolated-sectors}
\end{equation}

Because $u$ and $v$ vary independently in $\mathcal H_I$ and
$\mathcal H_J$, respectively, the first three types span
$\mathcal H_J\otimes\operatorname{span}\{\ket1\}$,
$\operatorname{span}\{\ket1\}\otimes\mathcal H_J$, and
$\mathcal H_J\otimes\mathcal H_I$, respectively. To separate the last two sectors, replace $u$ by $e^{i\phi}u$. The last
vector in Eq.~\eqref{eq:isolated-sectors} becomes
\begin{equation}
 e^{i\phi}\ket u\otimes\ket{\overline v}
 +\ket v\otimes\ket{\overline v}.
\end{equation}
Its zeroth and first Fourier coefficients in $\phi$ isolate
$\ket v\otimes\ket{\overline v}$ and
$\ket u\otimes\ket{\overline v}$, respectively. The latter vectors span
$\mathcal H_I\otimes\mathcal H_J$. Finally, the polarization identity gives
\begin{align}
\ket v\otimes\ket{\overline q}
&=\frac14\sum_{m=0}^{3}i^m
 (\ket v+i^m\ket q)\otimes
 (\ket{\overline v}+i^{-m}\ket{\overline q}),
\end{align}
so the complex span of
$\{\ket v\otimes\ket{\overline v}:v\in\mathcal H_J\}$ is
$\mathcal H_J\otimes\mathcal H_J$. 

We have therefore obtained all four sectors
\begin{equation}
 \mathcal K\otimes\mathcal K,
 \quad
 \mathcal K\otimes\mathcal H_J,
 \quad
 \mathcal H_J\otimes\mathcal K,
 \quad
 \mathcal H_J\otimes\mathcal H_J.
\end{equation}
Their direct sum is $\mathbb C^d\otimes\mathbb C^d$. Thus the zero product
vectors of $\widehat W^{(a)}_{d,r}$ span the full bipartite space. The spanning
property implies optimality \cite{Lewenstein2000}.

Finally, $D^{(a)}_{d,r}\ge0$. Therefore, for every positive operator $\rho$,
\begin{equation}
 \Tr\!\left(\widehat W^{(a)}_{d,r}\rho\right)
 =\Tr\!\left(W^{\max,(a)}_{d,r}\rho\right)
 -\Tr\!\left(D^{(a)}_{d,r}\rho\right)
 \le \Tr\!\left(W^{\max,(a)}_{d,r}\rho\right).
\label{eq:refinement-monotonicity-proof}
\end{equation}
Choose any embedded-qutrit state $\rho_{s,c}$ from
Eq.~\eqref{rsc} with negative expectation at
$w=z=2a$. Since
\begin{equation}
 \SN(\rho_{s,c})\le2,
 \qquad
 \SN(\rho_{s,c}^\Gamma)\le2,
\end{equation}
Eq.~\eqref{eq:refinement-monotonicity-proof} shows that the same state has
negative expectation on $\widehat W^{(a)}_{d,r}$. The certificate
Eq.~\eqref{eq:atomicity-certificate} therefore proves that the refined witness
is atomic, and hence indecomposable.
\end{proof}

It remains to compare the PPT states detected before and after the refinement.
\begin{remark}[Enlarged PPT-detection set]
\label{rem:refined-detection}
For every PPT state $\rho$,
Eq.~\eqref{eq:refinement-monotonicity-proof} holds. Hence the set of PPT states detected by the refinement contains the set
detected by the original endpoint witness, and it is strictly enlarged
whenever a PPT state satisfies
\begin{equation}
 0\le\Tr(W^{\max,(a)}_{d,r}\rho)<\Tr(D^{(a)}_{d,r}\rho).
\end{equation}
For the particular edge states on the full space
$R_{x,y}^{(d,r)}$ in Eq.~\eqref{eq:full-ppt}, however, the subtraction is
orthogonal to every rank-one sector in their displayed decomposition:
\begin{equation}
 \Tr(D^{(a)}_{d,r}R_{x,y}^{(d,r)})=0.
\end{equation}
Indeed, the vectors $\ket{1i}+\ket{i1}$ are orthogonal to
$\ket{1i}-\ket{i1}$, while $\ket{i\ell}-\ket{\ell i}$ is orthogonal to
$\ket{i\ell}+\ket{\ell i}$. Therefore the ellipse in
Eq.~\eqref{eq:detected-ellipse}, specialized to $w=z=2a$, is unchanged for
this analytic edge-state family. The gain in detection power concerns PPT
states having support on the symmetric $1$--$I$ sectors or the antisymmetric
$I$--$I$ sectors selected by $D^{(a)}_{d,r}$ of
\eqref{eq:refinement-subtraction}. 
\end{remark}

The inclusion is strict. To see this explicitly, choose the normalized edge state
\begin{equation}
 \sigma:=\frac 17 R^{(d,r)}_{1/r,\,1/p},
\end{equation}
for which
\begin{equation}
 \Tr(W^{\max,(a)}_{d,r}\sigma)=-\frac a7,
 \qquad
 \Tr(D^{(a)}_{d,r}\sigma)=0.
\end{equation}
Fix any $i\in I$ and let
$\nu _i:=\ket{1i}\!\bra{1i}$. Then $\nu _i$ is separable and
\begin{equation}
 \Tr(W^{\max,(a)}_{d,r}\nu_i)=a,
 \qquad
 \Tr(D^{(a)}_{d,r}\nu_i)=a.
\end{equation}
Therefore the PPT state
\begin{equation}
 \rho_\varepsilon:=(1-\varepsilon)\sigma+\varepsilon\nu_i
\end{equation}
satisfies
\begin{align}
 \Tr(W^{\max,(a)}_{d,r}\rho_\varepsilon)
 &=\frac a7(8\varepsilon-1),\\
 \Tr(\widehat W^{(a)}_{d,r}\rho_\varepsilon)
 &=-\frac a7(1-\varepsilon).
\end{align}
Hence, for $1/8<\varepsilon<1$, the original maximally coherent witness does
not detect $\rho_\varepsilon$, whereas the refined witness does. This also
proves directly that the enlargement of the PPT-detection set is strict.

\section{Conclusion}\label{sec:conclusion}
We have introduced a family of positive maps on qudits in which
independent exchange coherences compete with a collective diagonal
coherence. Despite its arbitrary local dimension, the family admits a
 complete analytic phase structure. The positivity region is universal, the
complete- and higher-positivity thresholds retain the channel multiplicities,
and the exact decomposable--atomic boundary is the circle
\begin{equation}
 (w-a)_+^2+(z-a)_+^2=a^2
\end{equation}
in scale-free coordinates. The cancellation of the multiplicities from this
boundary is one of the central structural features of the
model. 

The exact classification follows from complementary state- and map-side
constructions. An explicit representation $W=P+Q^\Gamma$ proves
decomposability on and inside the circular boundary, while a matched family
of PPT operators gives a negative expectation outside it. For positive state
parameters, these operators are PPT edge states. Their detection region is an
ellipse in the original variables and becomes a disk after a collective
rescaling. An embedded-qutrit certificate further proves that
indecomposability and atomicity coincide for every block-positive mixed member
of the family.

The higher-dimensional extension is not a passive embedding of the qutrit
construction. The collective multiplicity creates distinct
$k$-positivity thresholds, but the purely collective endpoint remains
decomposable throughout its block-positive interval. Thus the atomic phase
requires the coexistence, and competition, of the two coherence patterns. At
maximal coherence, a positive subtraction yields a witness with the
spanning property; the refined witness is optimal, remains atomic, and
strictly enlarges the set of detected PPT states.

These results provide an exact benchmark for numerical and
convex-optimization methods for positive maps and entanglement witnesses.
Natural extensions include additional coherence classes, nonuniform diagonal
weights, and more general sparse coupling graphs. Such models may reveal which
features found here, in particular the universal quadratic boundary and the
multiplicity cancellation, persist beyond the two-sector architecture.


\section*{Author contributions and artificial-intelligence disclosure}
The author conceived the project, developed the constructions, proved the
results, prepared the figures, and wrote the manuscript. GPT-5.6 Sol
was used for language polishing, formatting assistance, improving the
clarity of selected passages, and assisting with the generation and
revision of plotting code for the figures. No scientific data or
mathematical results were generated by the AI tool. The author
independently reviewed and validated all mathematical arguments,
plotting code, figure contents, and final wording.

\end{document}